\documentclass[11pt]{article}
\pdfoutput=1
\usepackage{jheppub}
\usepackage{graphicx}
\usepackage{amssymb, amsmath, amssymb}
\usepackage{slashed}
\usepackage{hyperref}
\usepackage{caption}
\usepackage{xcolor}
\usepackage{dsfont}
\usepackage{verbatim}
\usepackage{subfig}
\usepackage{float}

\newcommand\beq{\begin{equation}}
\newcommand\eeq{\end{equation}}
\newcommand\be{\begin{equation}}
\newcommand\ee{\end{equation}}

\title{Spectral Form Factor in Sparse SYK Models}

\preprint{UTTG-06-2022}

\author{Elena C\'{a}ceres, }
\author{Anderson Misobuchi, }
\author{Amir Raz}

\affiliation{University of Texas, Austin, Physics Department, Austin TX 78712, USA}

\emailAdd{elenac@utexas.edu, anderson.misobuchi@utexas.edu, araz@utexas.edu}

\abstract{We  investigate the spectral form factor  of the  sparse Sachdev-Ye-Kitaev model. We use numerical methods to establish that at  intermediate  times  the connected part of the spectral form factor is the dominant one. These connected contributions arise from  fluctuations around the disconnected geometry, not from a new saddle point. A similar effect was previously conjectured  in SYK but required a value of $N$ out of reach of current numerical simulations. }

\begin{document}
\maketitle

\section{Introduction}

One of the hurdles in understanding foundational aspects of holography is the scarcity of solvable strongly coupled quantum models that exhibit black hole physics. The Sachdev-Ye-Kitaev model (SYK) \cite{Kitaev_talk, MaldecenaStanford} is one such model. The SYK model is a quantum mechanical system of $N$ Majorana fermions coupled by a random all-to-all $q$-body interaction. In the large $N$ and low-energy limit, an approximate conformal symmetry emerges and the system saturates the chaos bound \cite{Chaos}. This unique combination of properties make it a remarkable toy model for a holographic dual of a two-dimensional black hole. The SYK model has spurred many developments in strongly correlated systems and in holography.

One interesting feature of SYK is that it  exhibits chaos. 
The Hamiltonian of a chaotic system, when considering small energy windows where the density of states is constant, is believed to resemble a random matrix. This insight due to Wigner \cite{Wigner}, implies that by studying statistical properties of random matrices subject to the symmetries of the Hamiltonian, we can understand the  statistical properties of energy levels and eigenstates of the system. One observable that captures the statistics of energy levels is the spectral form factor (SFF). 

In \cite{Cotler:2016fpe} the authors showed that the late time behavior of the spectral form factor in the all-to-all SYK is  governed by random matrix theory (RMT) just as expected for a chaotic system. At early times, the SFF decreases; this behavior is referred to as ``the slope". After certain time $T_d$ it reaches a minimum or dip. After $T_d$  the spectral form factor grows linearly along a ``ramp'',  until it saturates and  reaches a plateau at time $T_p.$ Beyond $T_p$ the value of the SFF is constant. Unlike the early time behavior, the behavior after $T_d$ is believed to be universal, valid for a large class of quantum mechanical interacting systems and has received much attention in the literature. 
In \cite{Cotler:2016fpe} the authors argued that before the dip the disconnected  part of the spectral form factor dominates. This matches the holographic intuition where at early times we expect the gravitational path integral to be dominated by two disconnected geometries. At late times, after the dip, the connected contributions are the dominant ones. 


Recently, in  \cite{Berkooz2021} the authors showed that at intermediate  time scales there are modifications to the connected
components that come from global fluctuations of the spectrum. These fluctuations were conjectured to be dominant at intermediate  times for sufficiently large $N$.  However, to numerically observe this behavior in the all-to-all would require very large values  of $N$ out reach of current numerical simulations.

In this work we focus on the intermediate time behavior of the spectral form factor in the \emph{sparse SYK} which allows for efficient numerical simulations. The sparse SYK model was proposed by Xu, Susskind, Su and Swingle (XSSS) \cite{xu2020}, see also \cite{Garcia2021}. In the sparse SYK is  the interactions grow linearly with $N$, as opposed to $N^q$, this leads to a significant reduction in computational complexity. It has been shown that, for appropriate sparsity,  the sparse SYK  presents the same features of the original all-to-all SYK model such as fast scrambling and maximal chaos at low temperatures. 

We investigate numerically the SFF of the sparse SYK model. We establish the existence of an intermediate time regime, before the dip, where the connected part dominates the SFF. The nature of these connected contributions is different form the ones at late times. 

At late times, the contribution of other saddle points that connect the two replicas  dominate. This  can be interpreted, holographically, as wormholes joining the geometries. The intermediate time contributions we find in this work, do not arise from a new saddle point, they  are due to fluctuations around the naive, disconnected, saddle geometry. Thus, it's interpretation from the holographic point of view is not clear but it indicates that in the large, but finite, $N$ limit, even two disconnected geometries do not fully factorize.


\section{The sparse SYK model}

We recall that the standard Sachdev-Ye-Kitaev (SYK) model consists of a system of $N$ Majorana fermions with $q$-body all-to-all random interactions, whose Hamiltonian is given by
\begin{equation} \label{eq:hamiltonian_all}
    H_\text{all-to-all} = \sum_{1\leq j_1<\ldots<j_q\leq N}J_{j_1\ldots j_q}\chi^{j_1}\ldots\chi^{j_q},
\end{equation}
where $\chi^j$, $j=1,\ldots, N$ denote the Majorana fermions, and the couplings $J_{j_1\ldots j_q}$ are sampled from a normal distribution with zero mean and variance
\begin{equation}
	\langle \left(J_{j_1\ldots j_q}\right)^2\rangle_\text{all-to-all} = \frac{(q-1)!J^2}{N^{q-1}}.
\end{equation}
The parameter $J$ above sets the typical energy scale. The advantage of this model is that it can be solved analytically in the large $N$ limit. On the other hand, the study of this model at finite $N$ relies on numerical simulations that quickly become intractable as we increase the number of fermions due to the increasing dimension of the Hilbert space and also due to the all-to-all nature of this model, in which the total number of terms grows as $\binom{N}{q}\sim N^q$. A variation of the SYK model, dubbed sparse SYK \cite{xu2020}, has been proposed as a more computationally tractable system that preserves the same features of the all-to-all SYK model, such as the quantum chaotic behavior \cite{xu2020, Garcia2021} and its connection to traversable wormholes \cite{caceres2021}. The sparse SYK model displays these properties with a Hamiltonian that contains only $\mathcal{O}(N)$ interaction terms.

The Hamiltonian of the sparse SYK model is
\begin{align} \label{eq:H_single}
	H & = i^{q/2}\sum_{j_1<\ldots <j_q}J_{j_1\ldots j_q}x_{j_1\ldots j_q}\chi^{j_1}\ldots\chi^{j_q}.
\end{align}
where we introduce the parameter $x_{j_1\ldots j_q}$ that can be either 0 or 1, such that different choices of $x_{j_1\ldots j_q}$ lead to different sparse models. In the literature, two implementations of sparsity have been considered. In the \textit{random pruning} approach, one chooses $x_{j_1\ldots j_q}=1$ with probability $p$ and $x_{j_1\ldots j_q}=0$ with probability $1-p$. Another approach is to consider \textit{regular hypergraphs}, in which the Majorana fermions are identified as vertices of a hypergraph whose hyperedges contain $q$ vertices. The regularity condition imposes that each fermion appears exactly the same number of times in the list of hyperedges. The hypergraph is also uniform, meaning that all hyperedges contain the same number, $q$, of vertices.

The random couplings associated to the sparse SYK model are still sampled from a Gaussian distribution of zero mean, but its variance is modified to be
\begin{equation} 
	\langle \left(J_{j_1\ldots j_q}\right)^2\rangle = \frac{(q-1)!J^2}{pN^{q-1}} = \frac{2^{q-1}\mathcal{J}^2(q-1)!}{pqN^{q-1}}
\end{equation}
In the regular hypergraph construction, the amount of sparseness is controlled by the parameter
\begin{equation} \label{eq:k}
    k \equiv \frac{p}{N}\binom{N}{q},
\end{equation}
which makes the Hamiltonian to be a sum of exactly $k\,N$ interaction terms.

\subsection{Analytical spectral density}

The spectrum of the all-to-all SYK model at finite $N$ and $q$ is very well approximated by a $Q$--Gaussian distribution \cite{Garc_a_Garc_a_2017,Garc_a_Garc_a_2018_2}, and the spectrum converges to this distribution in the double scaled limit of taking $N,q \rightarrow \infty$ but keeping $N/q^2$ fixed \cite{Erdos14,feng2018spectrum,Micha2018,Berkooz_2019}. This $Q$--Gaussian distribution is 
\begin{equation} \label{eq:spec_ana}
    \rho_\eta(\theta) = \frac{1}{2\pi}\left(\eta,e^{\pm2i\theta};\eta \right)_{\infty},
\end{equation}
where $(a;q)_{n}$ is the $Q$-Pochammer symbol
\begin{equation}
    (a;q)_{n} = \prod_{j=0}^{n-1}\left(1-aq^j \right), 
\end{equation}
with the shorthand notation $(a,b;q)_n \equiv (a;q)_n (b;q)_n$. 

In \eqref{eq:spec_ana} the angle $\theta\in[0,\pi]$ is related to the energies by 
\begin{equation}
    E(\theta) = \frac{2 {\cal{J}} \cos(\theta) }{\sqrt{1-\eta}},
\end{equation}
and the parameter $\eta$ depends on $q$ and $N$, and is given by \cite{Garc_a_Garc_a_2017}
\begin{equation}
    \eta(q,N) = {\binom{N}{q}}^{-1} \sum_{j=0}^q (-1)^j \binom{q}{j} \binom{N-q}{q-j}. 
\end{equation}

\begin{figure}
    \centering
    \includegraphics[width=\linewidth]{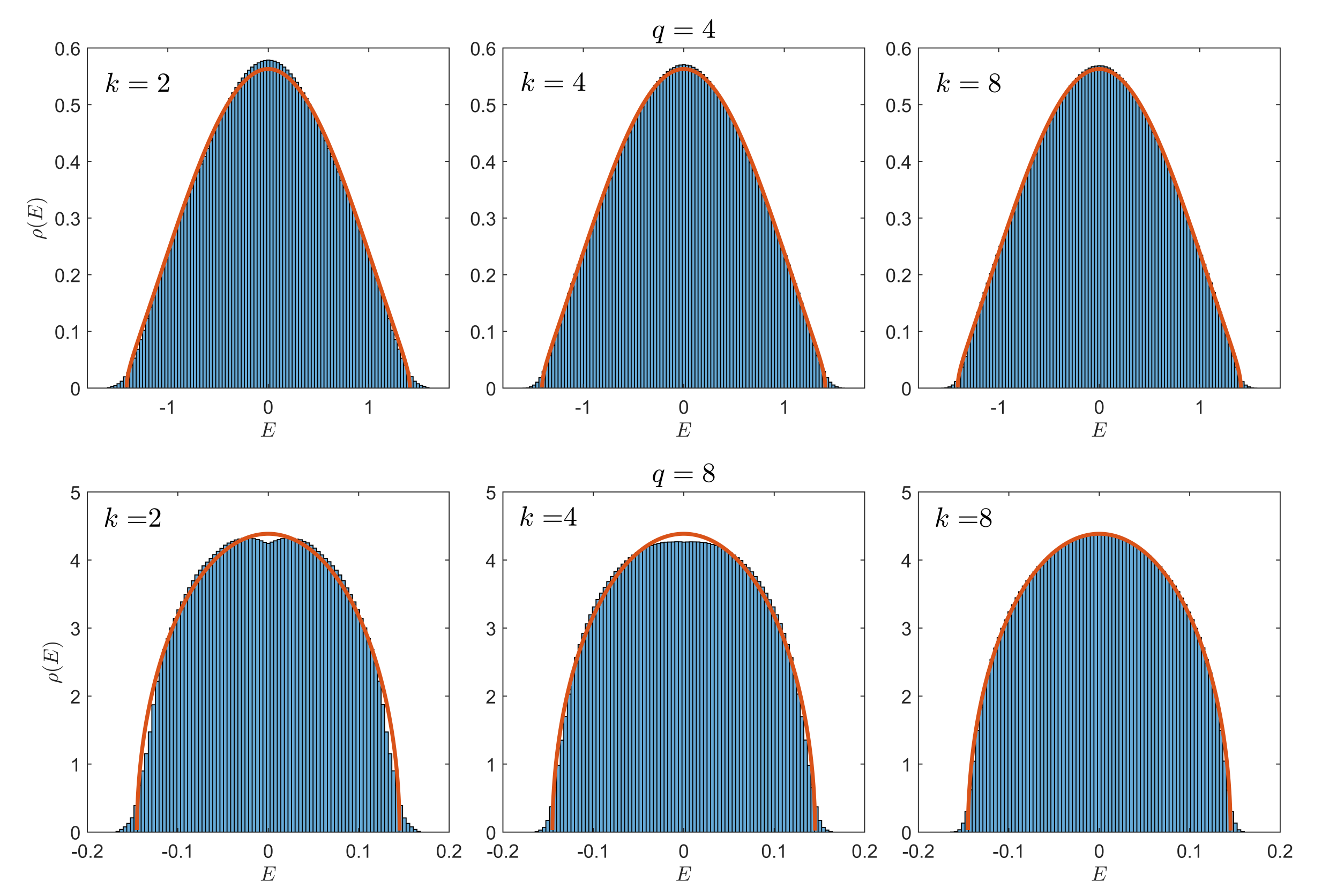}
    \caption{Histograms of the spectral density of the sparse SYK Hamiltonian for $N=30$ and various values of $k$. The spectral density is averaged over 100 disorder realizations. For comparison the solid red line is the $Q$--Gaussian approximation to the spectral density \eqref{eq:spec_ana}.}
    \label{fig:histograms_all}
\end{figure}

In the double scaled limit $\eta \rightarrow e^{-2q^2/N}$. Furthermore, the convergence of the analytical spectral density in this limit is independent of the precise distribution of the random couplings \cite{Erdos14,feng2018spectrum}, so the same exact result also holds for the sparse SYK model. Perhaps more surprisingly is that even at finite $N$ and $q$ the spectral density of the sparse SYK model is also well approximated by this same $Q$-Gaussian distribution, at least for large enough $N$ and $k$ \cite{Garcia2021}. One can add an additional correction to $\eta$, what \cite{Garcia2021} dubbed the ``renormalized" approximation, to make the agreement with spectral density even closer 
\begin{equation}
    \eta(q,N;k) = \eta(q,N) + \frac{3}{kN}. 
\end{equation}

Using this analytical spectral density the partition function can be approximated as
\begin{equation}
    \mathcal{Z}(\beta) = \text{tr}\left(e^{-\beta H} \right) \approx \int d\theta ~\rho_\eta(\theta) e^{-\beta E(\theta)} . 
\end{equation}
We will use this analytical approximation to give a general idea of the expected behavior of the spectral form factor when comparing to the numerical results.

To give a qualitative assessment of the validity of the $Q$--Gaussian approximation, we compare it to the average spectral density of the sparse SYK Hamiltonian computed using exact diagonalization in Fig.\ref{fig:histograms_all}. We see that for $q=4$ the spectrum has a $Q$-Gaussian shape with some small deviations from the analytical approximation, which is true even for $k=2$. On the other hand, the spectral density for $q=8$ deviates from the $Q$--Gaussian approximation at the center of the spectrum, where for small values $k\lesssim 4$ there is a noticeable dip. Similar observations on the validity of the $Q$--Gaussian approximation for $q=4$ were discussed in section III. B of  \cite{Garcia2021}.

\begin{figure}
    \centering
    \includegraphics[width=0.7\linewidth]{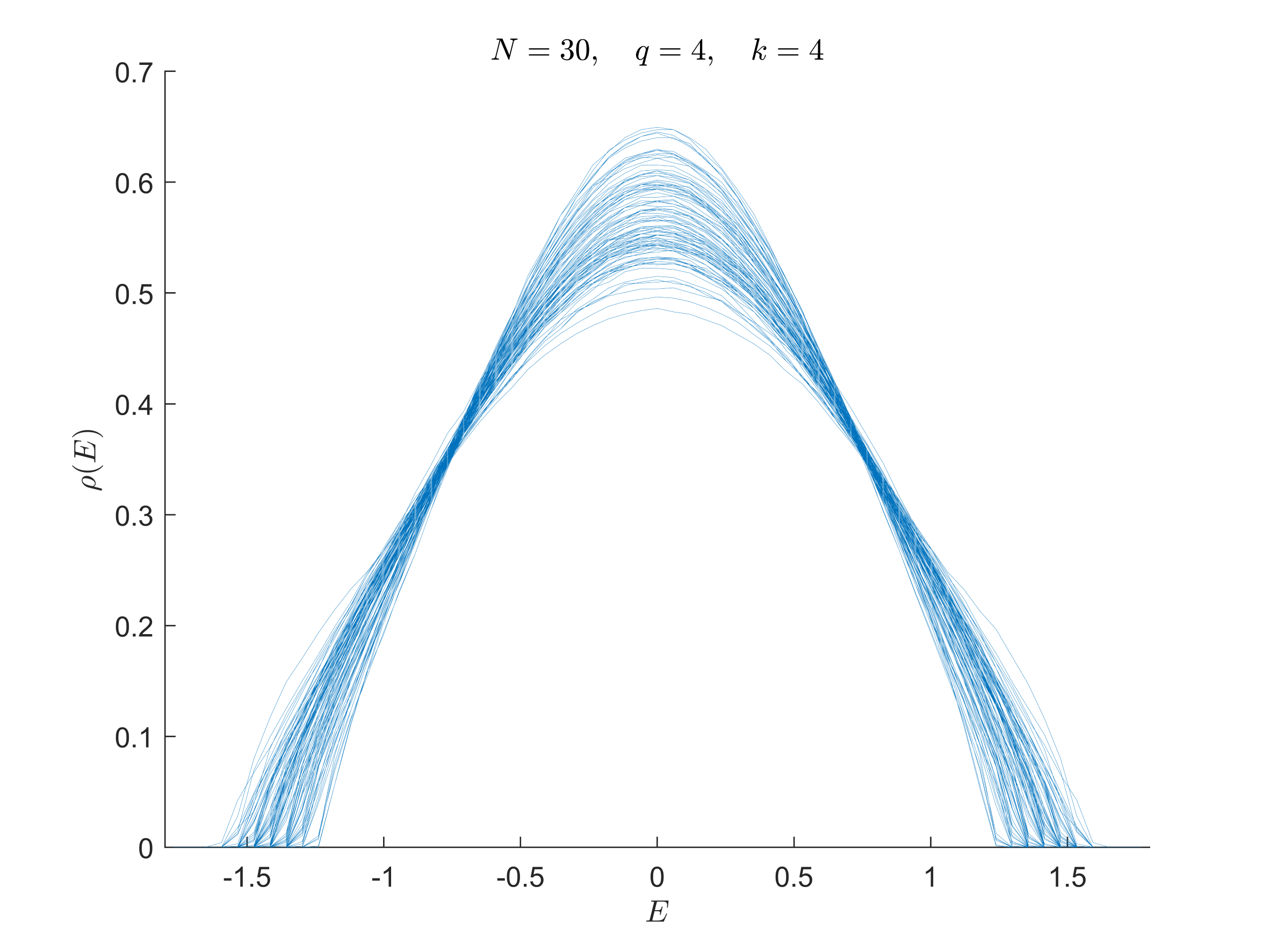}
    \caption{The individual spectral densities of the sparse SYK Hamiltonian for 100 different realization of the couplings.}
    \label{fig:histograms_q4_k4_spread}
\end{figure}

To understand how the spectral density deviates from the average in each realization of the couplings, we plotted the individual spectral densities for each of the 100 different realizations of couplings in Fig.\ref{fig:histograms_q4_k4_spread} for $N=30$, $q=4$, and $k = 4$. We see that the shape of the spectral density is similar for all the different disorder realizations, with the main deviation being a stretching or compressing of the spectrum; this corresponds to the dominant scaling mode described in \cite{Berkooz2021}. Additional plots of the spectral densities for other values of $k$ and $q$ and presented in the appendix.

\section{The spectral form factor}

The main focus of this paper is the spectral form factor (SFF) of the sparse SYK model. The SFF is defined by the two replica average
\begin{equation} \label{eq:specformOG}
    g(T;\beta) = \frac{\mathcal{Z}(\beta+iT,\beta-iT)}{[\mathcal{Z}(\beta)]^2}.
\end{equation}
The SFF contains the important information about the two replica energy correlations in disordered systems \cite{Br_zin_1997}. As such it has played a central role in the study of holography and quantum chaos
\cite{Cotler:2016fpe,saad2018,Liu_2018,del_Campo_2017,winer2021hydrodynamic}. For chaotic systems the late time SFF has a universal ramp and plateau behavior whose features are solely determined by the discrete symmetries (time reversal, charge, and parity) of the Hamiltonian \cite{Cotler:2016fpe,Liu_2018}. This universal behavior is described by a random matrix universality class, which typically fits into one of three RMT $\beta$-ensembles \cite{Liu_2018}.

From a holographic prospective, the spectral form factor describes the analytical continuation of the gravitational path integral over a space with two asymptotically AdS boundaries \cite{Cotler:2016fpe,saad2018,Saad:2019lba}. At early times we expect this gravitational path integral to be dominated by a geometry consisting of two disconnected disks, while at late times the connected geometries become the dominant configuration. These connected geometries are exponentially suppressed by factors of $e^{-S}$, where $S$ is the entropy.

These same dynamics happen in the SYK model. \cite{Cotler:2016fpe} showed that the SFF of the all-to-all SYK model behaves as the product of disconnected partition functions at early times, and follows the RMT prediction at exponentially late times (in $N$.)\footnote{We note that the late time RMT behavior was first observed by \cite{Garc_a_Garc_a_2016}, where they computed the Fourier transform of the SFF (i.e. the number variance.)} Later \cite{saad2018} showed that this late time behavior is due to the need to include new connected saddle points to the two replica path integral. These saddle points are indeed exponentially suppressed in $N$ (recall that the entropy of the SYK model is proportional to $N$.)

However the SFF of the SYK model also contains connected contributions that are only polynomially suppressed in $N$ \cite{Cotler:2016fpe,Berkooz2021}, the leading term being of order $N^{-q}$. These connected contribution do not arise from new saddle points, but rather from a perturbative expansion around the disconnected saddle point \cite{Berkooz2021}.\footnote{Essentially these contributions arise by computing the non-diagonal 1-loop determinant around the disconnected saddle order by order in $1/N$.} Though the existence of such contributions has been known for some time \cite{Cotler:2016fpe,feng2018spectrum2,Altland_2018,verbaarschot2019}, recently \cite{Berkooz2021} argued that this connected contribution should dominate the SFF of the all-to-all SYK model at intermediate times. These connected contributions have not been seen in the SFF of the all-to-all SYK model up to now due to finite $N$ constraints, though \cite{Berkooz2021} conjectured that $N=60$ would be enough to this new intermediate time behavior.

In  \cite{Berkooz2021}, the authors calculated the connected contributions for a general multi-trace thermal expectation value in the all-to-all SYK. In this work we are interested in the spectral form factor  \eqref{eq:specformOG} in the sparse SYK so let us first review the results of  \cite{Berkooz2021} for the double trace case that is the relevant one for the SFF.

Consider the thermal expectation value
\begin{equation}\label{eq:two_traces}
	\mathcal{Z}(\beta_1,\beta_2)= \left < \text{tr}(e^{-\beta_1 H)})\, \text{tr}(e^{-\beta_2 H})\right>_J\,,
	\end{equation}
where the sub index $J$ denotes the ensemble average over the random couplings. We can expand \eqref{eq:two_traces} as
\begin{align}
	\mathcal{Z}(\beta_1,\beta_2)&=  \sum_{k_1,k_2=0} ^\infty\left < \text{tr}(H^{k_1})\, \text{tr}(H^{k_2})\right>_J \frac{\beta_1^{k_1}}{{k_1}!} \frac{\beta_2^{k_2}}{{k_2}!} (-1)^{k_1+k_2}\\
&=\sum_{k_1,k_2=0}^\infty M(k_1,k_2) \frac{\beta_1^{k_1}}{{k_1}!} \frac{\beta_2^{k_2}}{{k_2}!} (-1)^{k_1+k_2} ,
\end{align}
where $M(k_1,k_2) $ denotes the joint moments of the distribution. We are interested in the leading order $1/N$ term in the connected part of these moments. The connected part can be obtained by subtracting all lower disconnected moments,
\begin{equation}
	M_c(k_1,k_2)=M(k_1,k_2) - M(k_1)M(k_2)\,.
\end{equation}
Therefore, the connected part of $	\mathcal{Z}(\beta_1,\beta_2)$ is simply given by the connected part of the moments,
\begin{align}\label{eq:Z2simple}
\mathcal{Z}_c(\beta_1,\beta_2) & = \sum_{k_1,k_2=0}^\infty M_c(k_1,k_2)\frac{\beta_1^{k_1}}{{k_1}!} \frac{\beta_2^{k_2}}{{k_2}!} (-1)^{k_1+k_2} ,
\end{align}
Furthermore, in the case of the all-to-all SYK, \cite{Berkooz2021} found that $M_c(k_1,k_2)$ is simply given by,
\begin{equation}\label{eq:moment2}
	M_c(k_1,k_2)=\frac{k_1 k_2}{2} \binom{N}{q}^{-1} M(k_1)M(k_2) + O(N^{-3q/2}).
\end{equation}
It is not clear how to interpret these connected contributions holographically as they are not associated to wormhole configurations in the bulk. Nevertheless, if these connected contributions dominate the SFF at some intermediate times then they must be important to the gravitational path integral. 

Thus, we can approximate the known contributions to the spectral form factor as 
\begin{equation} \label{eq:specform}
\begin{aligned}
    \mathcal{Z}(\beta+iT,\beta-iT) \approx |\mathcal{Z}(\beta + iT)|^2 & + 
    \frac{\epsilon}{2} \left(\beta^2+T^2\right) \left|\frac{\partial\mathcal{Z}(\beta')}{\partial(\beta')}\right|^2_{\beta' = \beta+iT} + \dots\\
    & +\int dE~e^{-2\beta E} \min\left\{ 4 \frac{T}{2\pi} , 2 \rho(E)\right\}.
    \end{aligned}
\end{equation}
The first term is the disconnected contribution, the second term is the leading connected piece (coming from \eqref{eq:Z2simple} and \eqref{eq:moment2}) and the third term is the RMT contribution given in \cite{Cotler:2016fpe}.\footnote{Here we assume the RMT contribution comes from a GUE ensemble, which is the case for $N\mod 8 = 2,6$ Other values of $N$ fall into the GSE or GOE universality classes, and the ramp behavior is more complex, and can be found in \cite{Liu_2018}.} The parameter $\epsilon$ is a small parameter convenient to organize higher order corrections (denoted by the dots in \eqref{eq:specform}), which for the all-to-all SYK is $\epsilon = {\binom{N}{q}}^{-1}$.

For the sparse SYK model, one can use the same analysis in \cite{Berkooz2021} to compute the leading connected contribution to the SFF assuming the random couplings are non-Gaussian. In this case the leading connected contribution only depends on the normalized forth moment of the couplings,\footnote{See Theorem 1 of \cite{feng2018spectrum2} for a rigorous mathematical derivation.} with
\begin{equation}
    \epsilon = \frac{1}{2}\binom{N}{q}^{-1}\left( \gamma-1\right),
\end{equation}
where $\gamma$ is the normalized fourth moment of the couplings
\begin{equation}
    \gamma = \frac{\left<\left( J_{j_1\ldots j_q}x_{j_1\ldots j_q} \right)^4\right> }{ \left<\left( J_{j_1\ldots j_q}x_{j_1\ldots j_q} \right)^2 \right>^2} .
\end{equation}
Thus even for the sparse SYK model the SFF still takes the form \eqref{eq:specform}, only with 
\begin{equation}\label{eq:epsilon}
    \epsilon = \frac{1}{2}\binom{N}{q}^{-1}\left( \frac{3}{p}-1\right) \approx \frac{3}{2kN}.
\end{equation}
Note that when $p\rightarrow 1$ we recover the all-to-all result, as expected.

Assuming $k = \mathcal{O}(1)$ we see that the suppression of the leading connected term in the sparse SYK model is only of order $1/N$, and so we expect this contribution to dominate at earlier times. Hence we may be able to see the intermediate time behavior conjectured in \cite{Berkooz2021} at manageable values of $N$'s for the sparse SYK model.

Finally, we note that many in the condensed matter (and high energy) community are primarily interested in the SFF to understand the spectral correlation. These correlations are easier to study if the energies are re-scaled so that the local density of states is a constant, in a process called spectral unfolding. The spectral unfolding is typically done by renormalizing the spectrum of each disorder realization independently\footnote{Occasionally in the literature the spectrum of the whole ensemble is unfolded together, rather than independently in each disorder realization. In this case the global spectral correlations will remain and our conclusions will still be valid.} (for example by using a smooth function that is fitted to the numerical data of each realization.) This local unfolding will remove the global correlations in the SFF and only leave the local correlations, i.e. the universal RMT part. This was seen in \cite{verbaarschot2019} for the all-to-all SYK model, and in \cite{Garcia2021} for the sparse SYK model. One can further focus only on the bulk of the spectrum by adding a Gaussian filter, as was done in \cite{verbaarschot2019,Garcia2021,untajs_2020}. Such a filter, in combination with local spectral unfolding,  removes any edge and global effects, and is particularly powerful in understanding the local eigenvalue statistics.

In this work we focus on the full SFF at finite temperature rather than the unfolded SFF as it has a particularly simple holographic interpretation, and contains the global fluctuations which we are particularly interested in. One can add a Gaussian filter to the SFF (or any other smooth filter,) and calculate the leading contribution to the SFF (without unfolding) using the same moment method described above, though it is not clear to us what the holographic interpretation of these observables is.


\section{Numerical results}

We evaluated the spectral form factor using exact diagonalization for $N=30$ Majorana fermions. Our numerical results are evaluate considering the regular hypergraph model. Although in principle the derivation of \eqref{eq:epsilon} assumes the random pruning definition of the sparse SYK model since it treats $x_{j_1,\ldots, j_q}$ as a random variable, we expect corrections to this result when one considers a regular hypergraph instead to the moments to be of an order of $1/N^{q-1}$ \cite{Garcia2021}. We have confirmed this numerically so we will only display the results for the regular hypergraph model.

\begin{figure}
    \centering
    \includegraphics[width=0.7\linewidth]{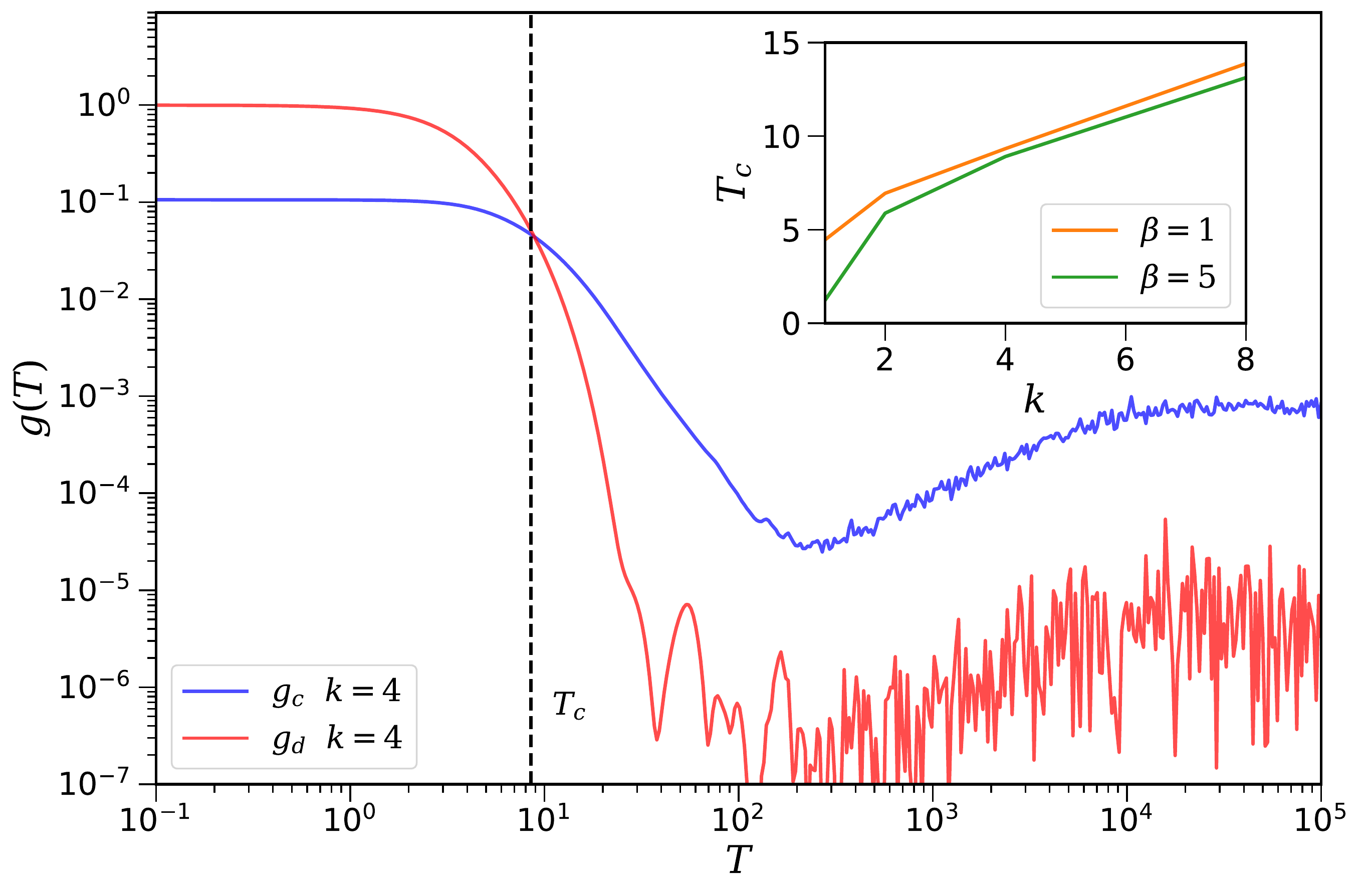}
    \caption{Main: Exchange of dominance in the spectral form factor of sparse SYK model, for $N=30$, $q=4$, $k=4$, $\beta = 5$, averaged over 100 disorder realizations. Inset: the critical time $T_c$ dependence on the sparsity $k$ and inverse temperature $\beta$.}
    \label{fig:tc}
\end{figure}

The average over the random couplings is taken by averaging the numerator and denominator separately, i.e.
\begin{align}
    g(T, \beta) & = \frac{\langle \mathcal{Z}(\beta, T)\,Z^*(\beta, T)\rangle_J}{\langle Z(\beta)\rangle_J^2}\,, \\
g_d(T, \beta) & = \frac{\langle \mathcal{Z}(\beta, T)\rangle_J \, \langle Z^*(\beta, T)\rangle_J}{\langle \mathcal{Z}(\beta)\rangle_J^2}\,, \\
g_c(T, \beta) & = g(T, \beta) - g_d(T, \beta)\,.
\end{align}

We find the existence of a transition at some time $t_c$ (before the dip) where the connected part of the spectral form factor begins to dominate over the disconnected part -- see Fig.\,\ref{fig:tc}. The transition time appears to increase linearly with the sparsity parameter $k$, and seems to slightly decrease when the inverse temperature $\beta$ is increased. 

\begin{figure}
    \centering
    \includegraphics[width=\linewidth]{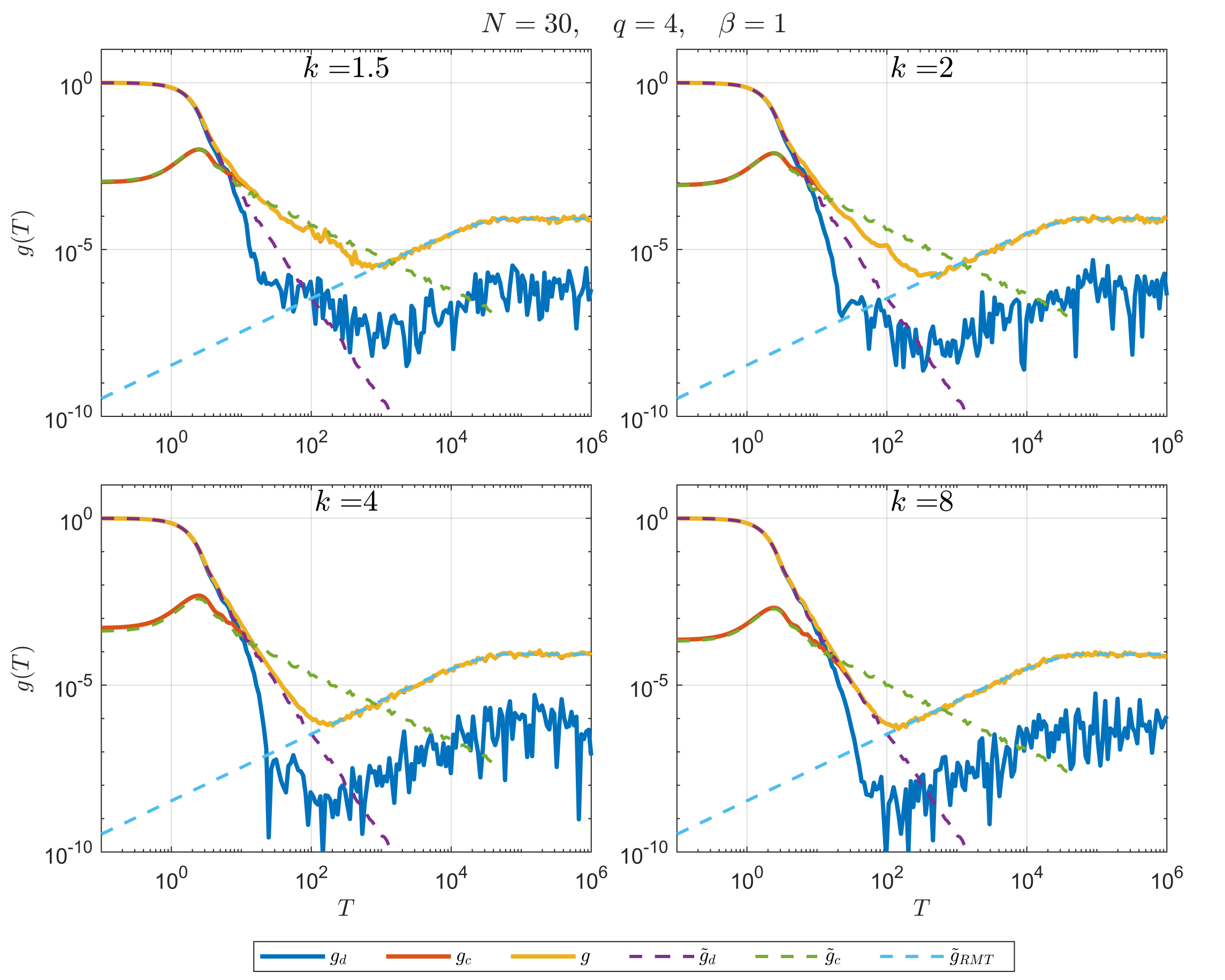}
    \caption{Full spectral form factor $g$, connected part $g_c$, and disconnected part $g_d$, obtained by exact diagonalization of the sparse SYK Hamiltonian define over a regular hypergraph, and averaged over 100 disorder realizations of the random couplings. These are compared to the analytical $Q$--Gaussian approximation $\tilde{g}_c$, $\tilde{g}_d$, and $\tilde{g}_\text{RMT}$.}
    \label{fig:SFF_q4_b1}
\end{figure}

We also find that the computed SFF from exact diagonalization is in agreement with the analytical formula \eqref{eq:specform} up to a timescale close to the point where the transition takes place and the agreement becomes significantly better as the sparsity parameter $k$ increases -- see Fig. \ref{fig:SFF_q4_b1}. This is true even for the small value of $k = 1.5$, which is somewhat surprising. After the exchange of dominance the connected SFF does not agree with the first order analytical approximation, at least for $k\gtrsim 2$. Somewhat surprisingly, the connected contribution seems to follow the disconnected approximation for large enough $k \gtrsim 4$. This behavior may be because the SFF is still dominated by the same saddle point, even though this saddle no longer factorizes into the two disconnected saddle points.  A similar behavior of the SFF occurs when $q=8$ and for other values of $\beta$; these additional plots of the SFF are given in the appendix.

\section{Discussion and holographic interpretation}

 We have observed an intermediate time transition in the SFF of the sparse SYK model, where the SFF is no longer dominated by the disconnected contribution. To understand how to interpret this transition in the dual gravitational description, it is illuminating to consider the path integral description of the SYK model where these connected contributions arise from a perturbative expansion around the disconnected saddle \footnote{See section 8 of \cite{Berkooz2021} for the full derivation.}. Thus this intermediate time regime is still described by a disconnected saddle geometry, but the connected fluctuations around the saddle are in fact important\footnote{This is similar to a double trace deformation of the boundary CFT \cite{Aharony:2001pa,Witten:2001ua,Berkooz:2002ug}.}. This raises a factorization problem in the dual holographic description which is stronger than the usual factorization problem in quantum gravity (à la Maldacena-Maoz \cite{MaldacenaMaoz}), as even two disconnected geometries do not factorize. 

 We may expect such connected contributions to be generic in models of holography, as these models are typically only understood perturbatively in some large parameter $N$, but have an exponential number states which goes like $e^{\# N}$. As a result the connected multiple boundary saddles are exponentially suppressed in $N$, and so are relevant only at exponentially large times. However the disconnected saddle is only understood holographically to leading order in $N$, and so the disconnected picture is only valid for times of order 1. At intermediate times of order $N$ (or more generally $N$ to some positive power), $1/N$ corrections become relevant. The $1/N$ corrections should generically contain any terms allowed by the symmetry, and so should contain disconnected contributions; for example via double trace deformations.

Our results suggest several directions to explore in the future:
\begin{itemize}
\item \textbf{ Holographic interpretation} \hfill\break
As mentioned above, the  intermediate  time dominant  contributions  presented here are connected fluctuations 
around the disconnected saddle geometry; They do not correspond to a new, connected saddle. Thus,  it is natural to ask what  is the  bulk interpretation of this regime.  In \cite{Berkooz2021}, the authors suggested that these connected contributions correspond to the existence of light  fields in the bulk with fluctuating boundary couplings.  It would be interesting to study this issue further.

\item \textbf{Other observables} \hfill\break
The connected contributions that dominate the SFF at time before the dip time, should also leave an imprint in other observables. We could consider for example, the new measure  of quantum state complexity defined by minimizing the spread of the wavefunction over all choices of basis proposed in \cite{Balasubramanian:2022tpr}.  In systems with chaotic behavior this notion of  complexity displays four
dynamical regimes: a  linear ramp, a peak and  a slope down to a plateau.  This behavior can be connected to the ramp -dip- plateau structure of the SFF.  Furthermore, the complexity slope,  arises from spectral rigidity just like the SFF ramp. It would be interesting to understand how the dominance of connected fluctuations studied here is reflected in this new measure  complexity. 

\item \textbf{Larger $N$ simulation} \hfill\break
Our numerical results relied on the exact diagonalization of the Hamiltonian which significantly constrains the values of $N$ that we can achieve. This was enough to demonstrate the exchange of dominance at intermediate times in the sparse SYK model, but the same cannot be said for the all-to-all SYK, where one expects at least $N=60$ to observe the same effect. Recent techniques such as Krylov subspace methods for time evolution have been employed to study quantum chaos in the SYK model for up to $N=60$ \cite{kobrin2020manybody}. However, we found that these techniques, when applied to the SFF, are susceptible to large errors so that it fails to correctly capture the correlation between eigenvalues. It would be desirable to refine or find an alternative algorithm such that the numerical simulation could be pushed to larger values of $N$, allowing us to verify the same statement for the all-to-all SYK.

\end{itemize}

\acknowledgments

We would like to thank Micha Berkooz, Vladimir Narovlansky and Antonio Garcia-Garcia for comments on the manuscript. We also thank Kristan Jensen for enlightening discussions. E.C. and A.M. are supported by National Science Foundation (NSF) Grant No. PHY-2112725. A.R. is supported by the U.S. Department of Energy under Grant No. DE-SC0022021 and a grant from the Simons Foundation (Grant 651440, AK). E.C also thanks the Aspen Center for Physics, supported by National Science Foundation grant PHY-1607611, where part of this work was performed.


\appendix

\section{Appendix: Extended numerical results}

\begin{figure}[h]
    \centering
    \includegraphics[width=\linewidth]{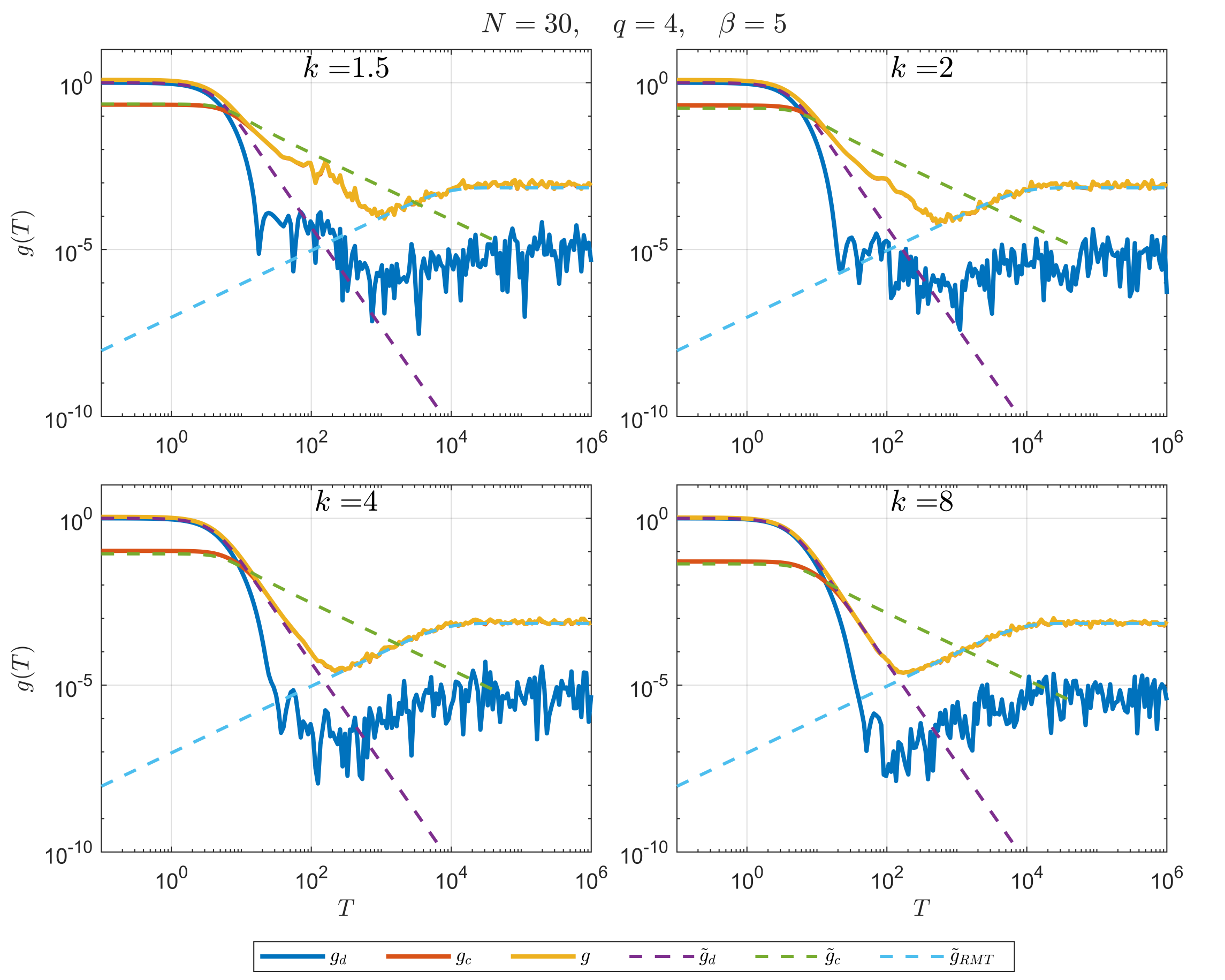}
    \caption{Full spectral form factor $g$, connected part $g_c$, and disconnected part $g_d$, obtained by exact diagonalization of the sparse SYK Hamiltonian define over a regular hypergraph, and averaged over 100 disorder realizations of the random couplings. These are compared to the analytical $Q$--Gaussian approximation $\tilde{g}_c$, $\tilde{g}_d$, and $\tilde{g}_\text{RMT}$.}
    \label{fig:SFF_q4_b5}
\end{figure}

\begin{figure}
    \centering
    \includegraphics[width=\linewidth]{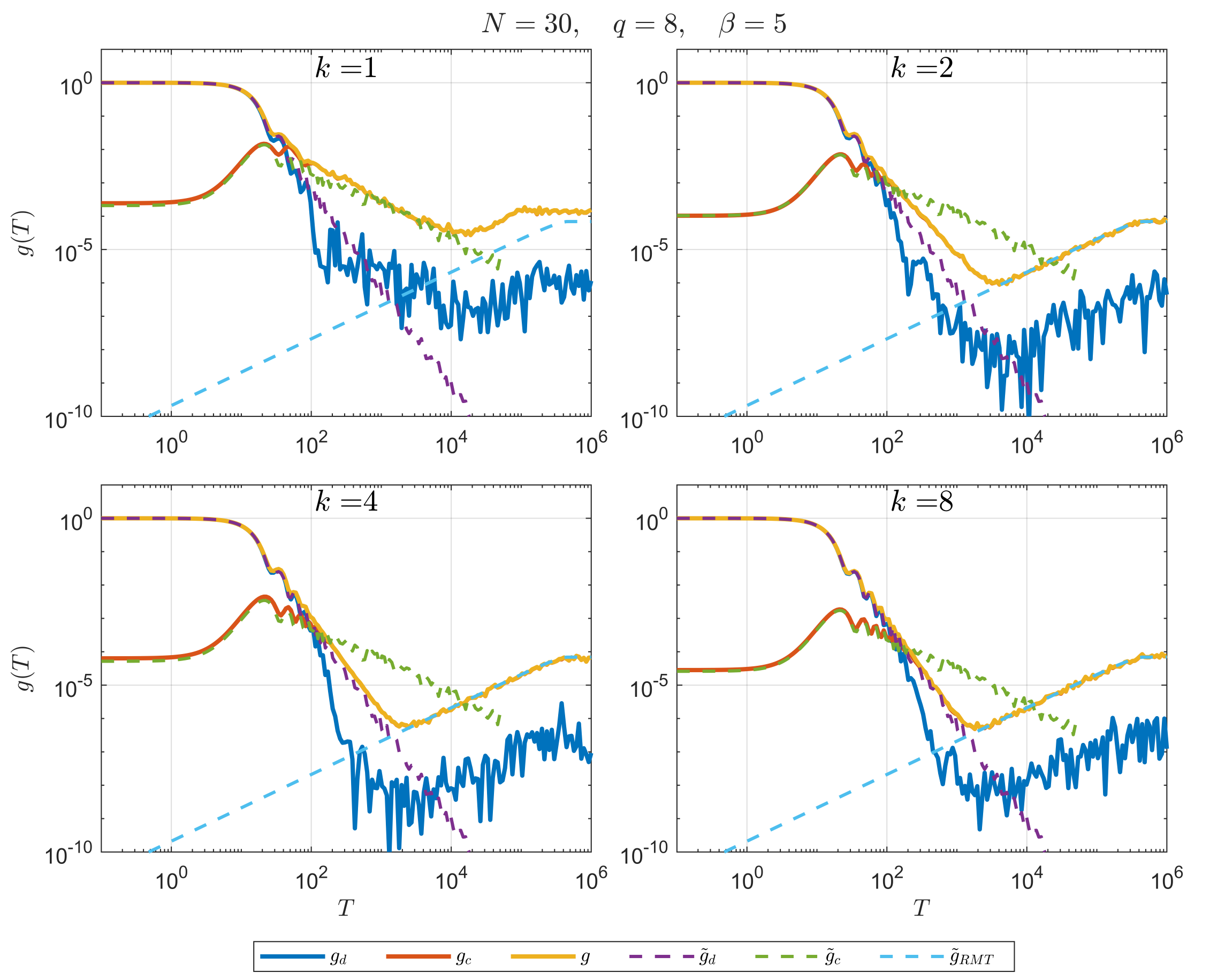}
    \caption{Full spectral form factor $g$, connected part $g_c$, and disconnected part $g_d$, obtained by exact diagonalization of the sparse SYK Hamiltonian define over a regular hypergraph, and averaged over 100 disorder realizations of the random couplings. These are compared to the analytical $Q$--Gaussian approximation $\tilde{g}_c$, $\tilde{g}_d$, and $\tilde{g}_\text{RMT}$.}
    \label{fig:SFF_q8_b5}
\end{figure}

\begin{figure}
    \centering
    \includegraphics[width=\linewidth]{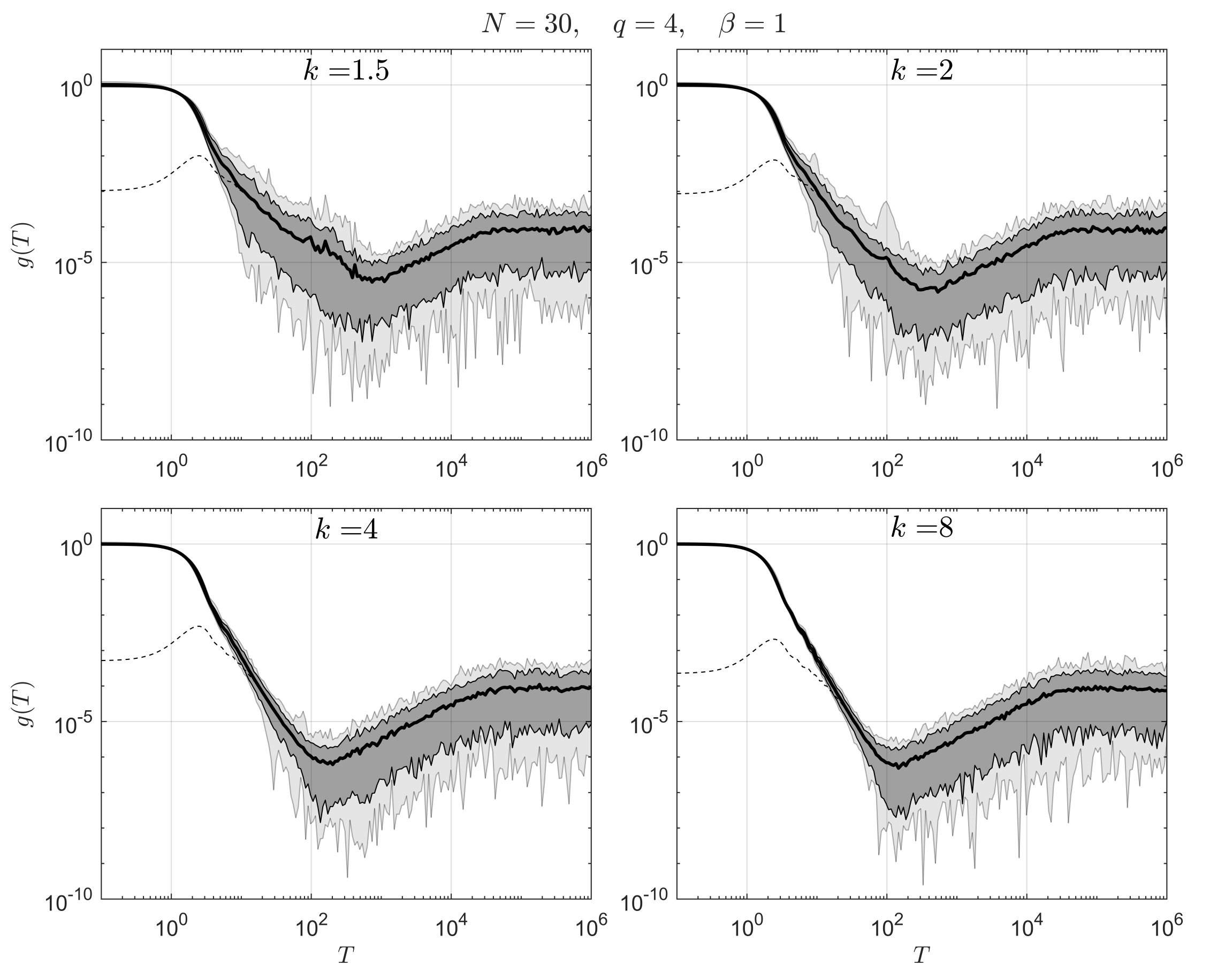}
    \caption{The spread of the spectral form factor over the 100 disorder realizations of the random couplings. The black line is the disorder averaged spectral form factor, the light gray area is the maximal spread of the SFFs, while the dark gray area contains 90\% of the disorder realizations. The disconnected part of the SFF is plotted in a dashed line to give some indication of time scales.}
    \label{fig:SFFspread_q4_b1}
\end{figure}

\begin{figure}
    \centering
    \includegraphics[width=\linewidth]{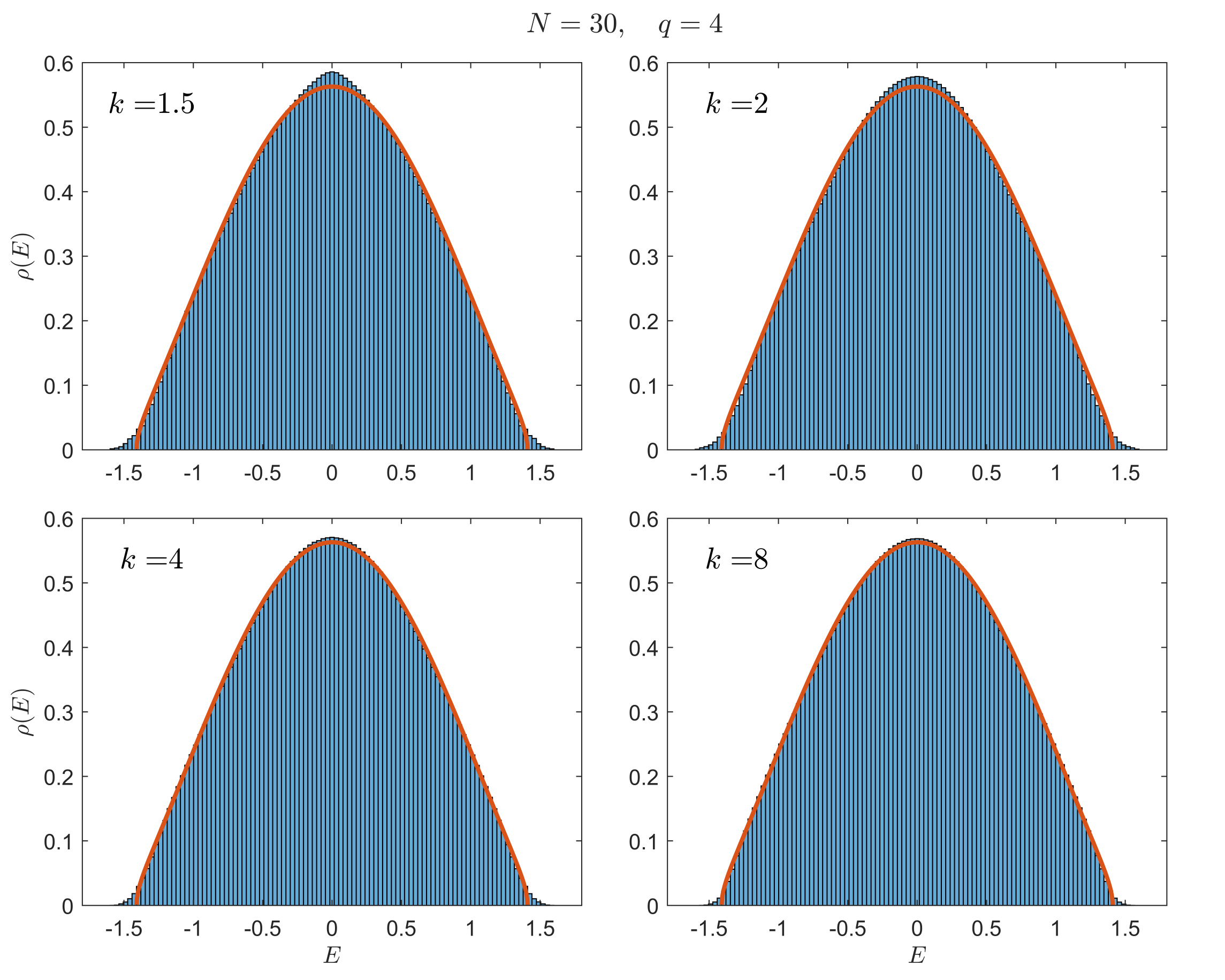}
    \caption{A histogram of the spectral density of the sparse SYK Hamiltonian averaged over 100 disorder realizations for various values of $k$. For comparison the solid red line is the $Q$--Gaussian approximation to the spectral density \eqref{eq:spec_ana}.}
    \label{fig:histograms_q4}
\end{figure}

\begin{figure}
    \centering
    \includegraphics[width=\linewidth]{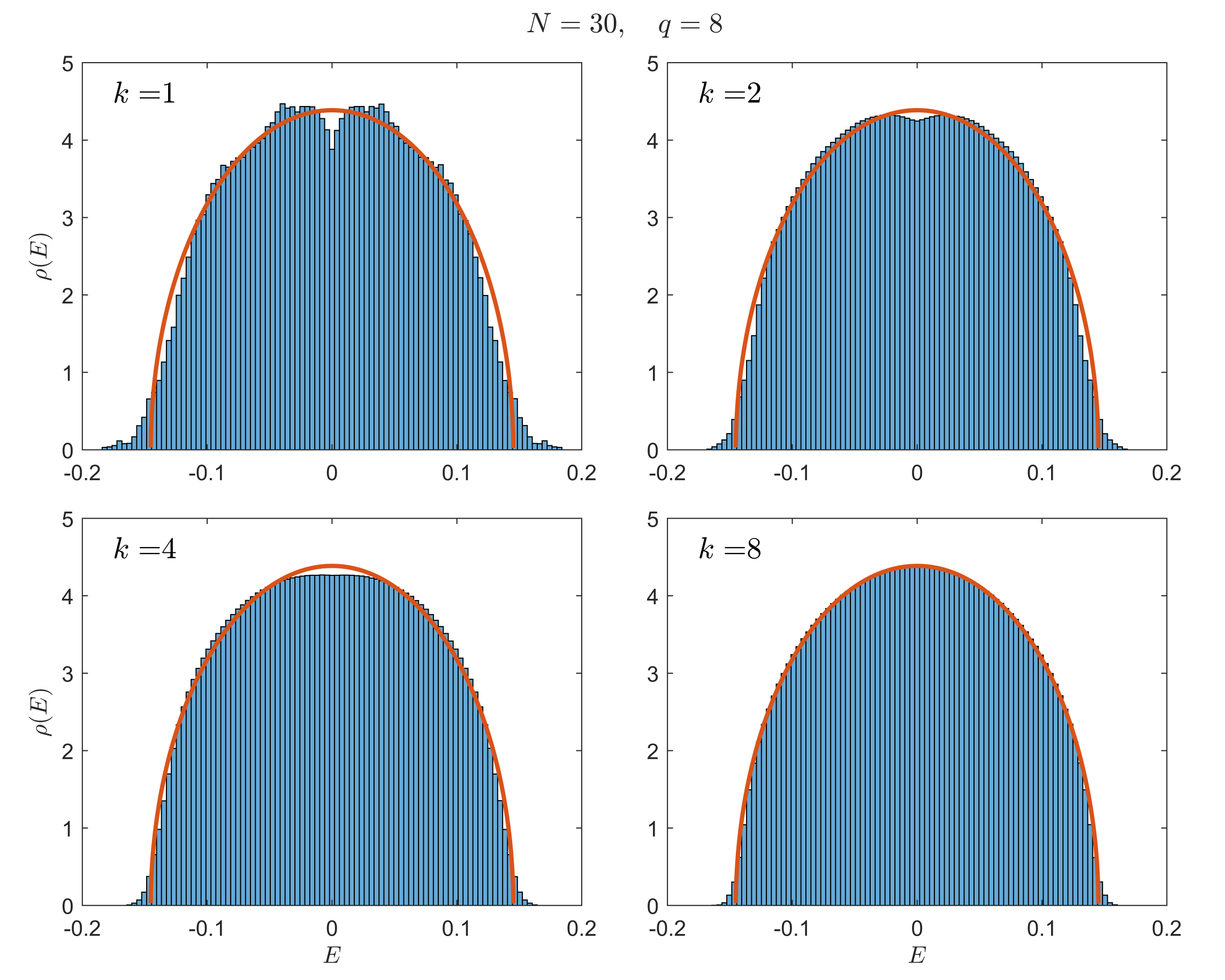}
    \caption{A histogram of the spectral density of the sparse SYK Hamiltonian averaged over 100 disorder realizations for various values of $k$. For comparison the solid red line is the $Q$--Gaussian approximation to the spectral density \eqref{eq:spec_ana}.}
    \label{fig:histograms_q8}
\end{figure}

\begin{figure}
    \centering
    \includegraphics[width=\linewidth]{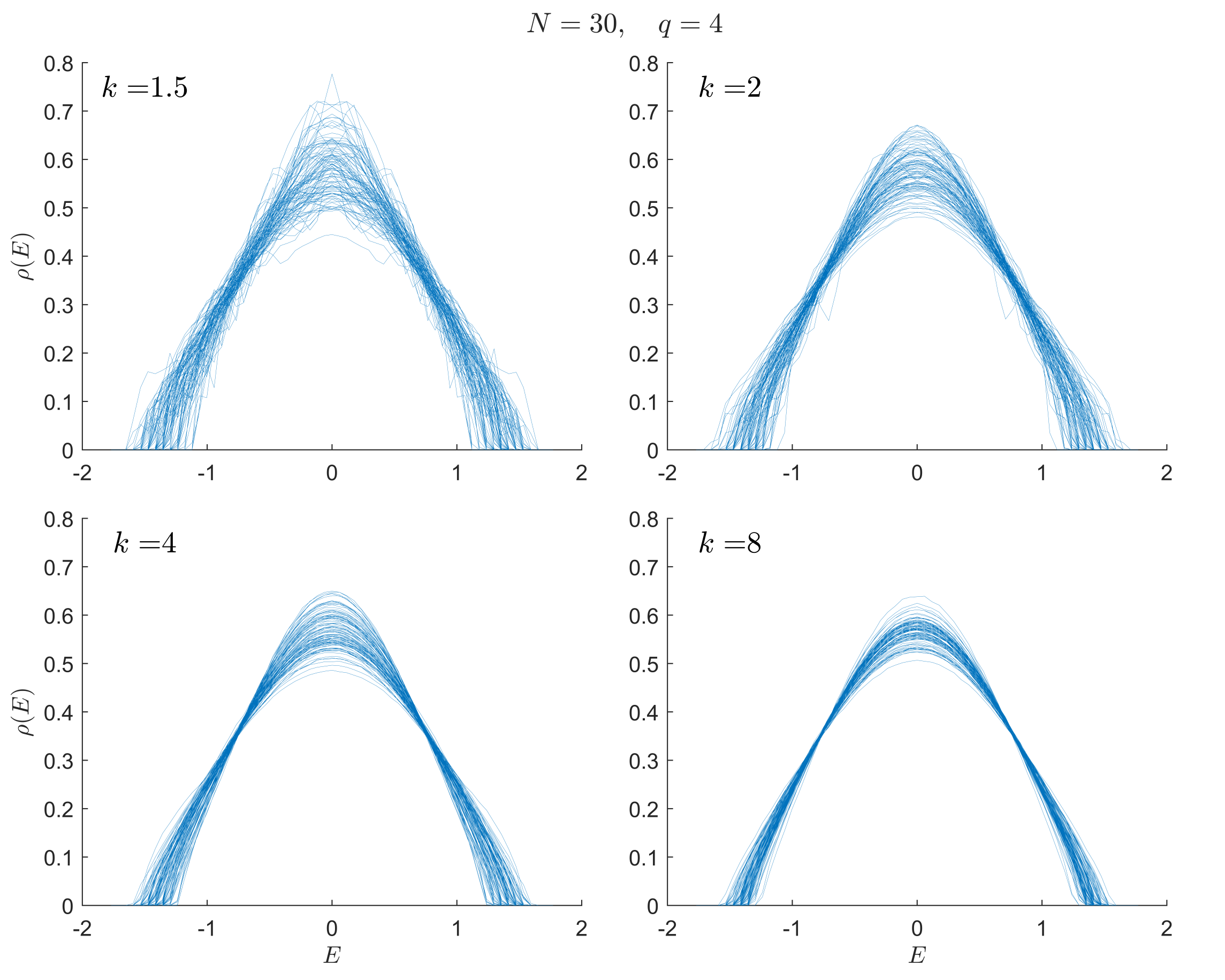}
    \caption{The spread of the spectral density of the sparse SYK Hamiltonian for 100 disorder realizations and for various values of $k$. }
    \label{fig:histograms_q4_spread}
\end{figure}

\begin{figure}
    \centering
    \includegraphics[width=\linewidth]{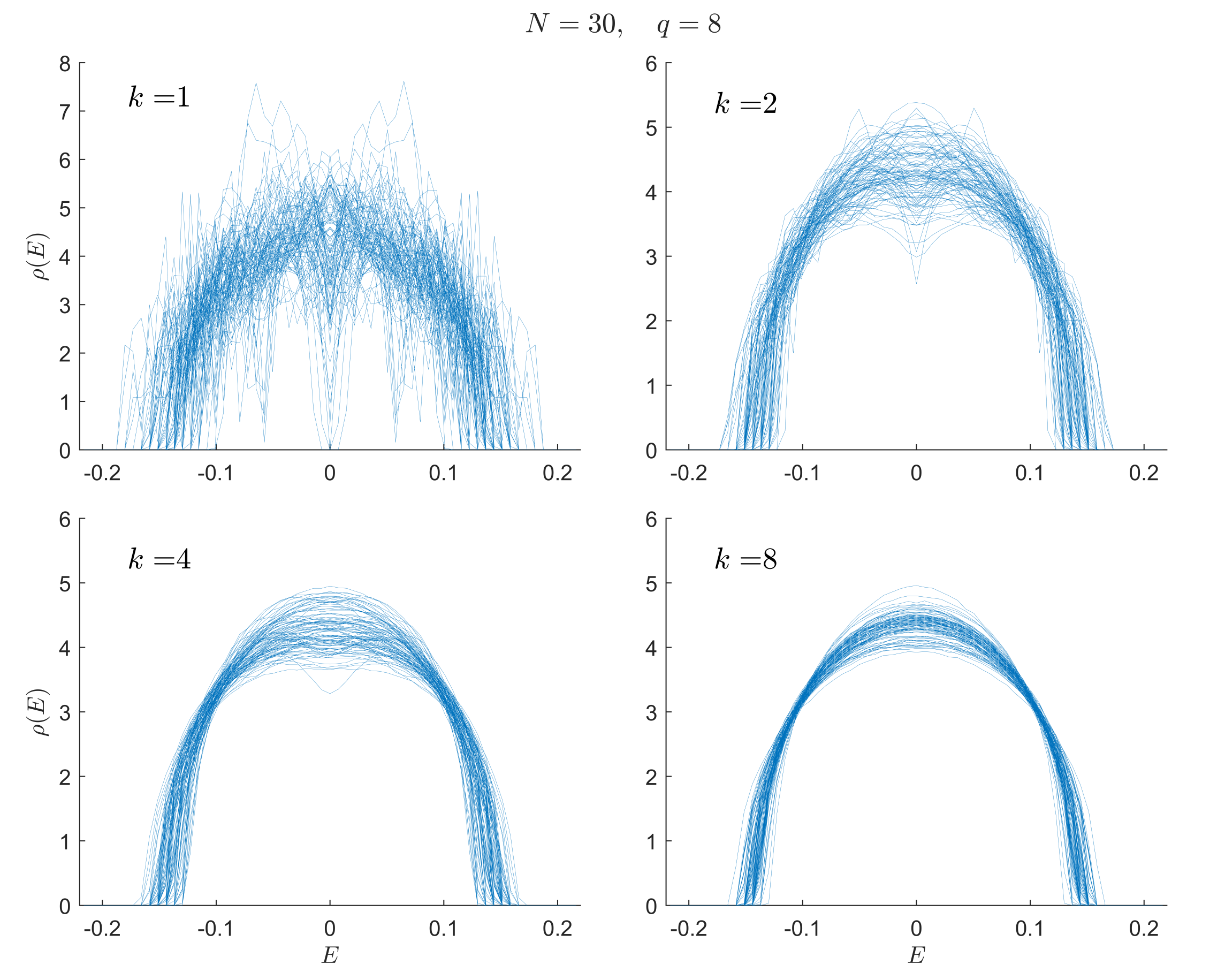}
    \caption{The spread of the spectral density of the sparse SYK Hamiltonian for 100 disorder realizations and for various values of $k$. }
    \label{fig:histograms_q8_spread}
\end{figure}

In this appendix we offer additional plots of the numerical results discussed in the text. In Fig.\ref{fig:SFF_q4_b5} and \ref{fig:SFF_q8_b5} we give additional plots of the SFF taking $\beta =5$ and $q=4$ and $q=8$. To give a qualitative picture of the variance of the SFF we plotted the region containing the individual SFF of all 100 disorder realizations, and the region containing 90 disorder realizations in Fig.\ref{fig:SFFspread_q4_b1}. In Fig.\ref{fig:histograms_q4} and \ref{fig:histograms_q8} we offer additional histograms of the average spectral density, and compare them to the Q--Gaussian approximation. Finally in Fig.\ref{fig:histograms_q4_spread} and \ref{fig:histograms_q8_spread} we present additional plots of the individual spectral densities for the different realization of the couplings.

\clearpage
\bibliographystyle{JHEP}
\bibliography{SFF_Sparse_SYK}

\providecommand{\href}[2]{#2}\begingroup\raggedright\begin{thebibliography}{10}

\bibitem{Kitaev_talk}
A.~Kitaev, \emph{{A simple model of quantum holography}}.
  \url{http://online.kitp.ucsb.edu/online/entangled15/kitaev/},
  \url{http://online.kitp.ucsb.edu/online/entangled15/kitaev2/}.

\bibitem{MaldecenaStanford}
J.~Maldacena and D.~Stanford, \emph{{Remarks on the Sachdev-Ye-Kitaev model}},
  \href{https://doi.org/10.1103/PhysRevD.94.106002}{\emph{Phys. Rev.}
  {\bfseries D94} (2016) 106002}
  [\href{https://arxiv.org/abs/1604.07818}{{\ttfamily 1604.07818}}].

\bibitem{Chaos}
J.~Maldacena, S.~H. Shenker and D.~Stanford, \emph{A bound on chaos},
  \href{https://doi.org/10.1007/JHEP08(2016)106}{\emph{Journal of High Energy
  Physics} {\bfseries 2016} (2016) 106}.

\bibitem{Wigner}
E.~P. Wigner, \emph{Characteristic vectors of bordered matrices with infinite
  dimensions}, {\emph{Annals of Mathematics} {\bfseries 62} (1955) 548}.

\bibitem{Cotler:2016fpe}
J.~S. Cotler, G.~Gur-Ari, M.~Hanada, J.~Polchinski, P.~Saad, S.~H. Shenker
  et~al., \emph{{Black Holes and Random Matrices}},
  \href{https://doi.org/10.1007/JHEP09(2018)002,
  10.1007/JHEP05(2017)118}{\emph{JHEP} {\bfseries 05} (2017) 118}
  [\href{https://arxiv.org/abs/1611.04650}{{\ttfamily 1611.04650}}].

\bibitem{Berkooz2021}
M.~Berkooz, N.~Brukner, V.~Narovlansky and A.~Raz, \emph{Multi-trace
  correlators in the syk model and non-geometric wormholes},
  \href{https://doi.org/10.1007/jhep09(2021)196}{\emph{Journal of High Energy
  Physics} {\bfseries 2021} (2021) }.

\bibitem{xu2020}
S.~Xu, L.~Susskind, Y.~Su and B.~Swingle, \emph{A sparse model of quantum
  holography},  2020.

\bibitem{Garcia2021}
A.~M. Garcia-Garcia, Y.~Jia, D.~Rosa and J.~J. Verbaarschot, \emph{Sparse
  sachdev-ye-kitaev model, quantum chaos, and gravity duals},
  \href{https://doi.org/10.1103/physrevd.103.106002}{\emph{Physical Review D}
  {\bfseries 103} (2021) }.

\bibitem{caceres2021}
E.~Caceres, A.~Misobuchi and R.~Pimentel, \emph{Sparse syk and traversable
  wormholes},  2021.

\bibitem{Garc_a_Garc_a_2017}
A.~M. García-García and J.~J. Verbaarschot, \emph{Analytical spectral density
  of the sachdev-ye-kitaev model at finite n},
  \href{https://doi.org/10.1103/physrevd.96.066012}{\emph{Physical Review D}
  {\bfseries 96} (2017) }.

\bibitem{Garc_a_Garc_a_2018_2}
A.~M. García-García, Y.~Jia and J.~J.~M. Verbaarschot, \emph{Exact moments of
  the sachdev-ye-kitaev model up to order 1/n2},
  \href{https://doi.org/10.1007/jhep04(2018)146}{\emph{Journal of High Energy
  Physics} {\bfseries 2018} (2018) }.

\bibitem{Erdos14}
L.~Erdős and D.~Schröder, \emph{{Phase Transition in the Density of States of
  Quantum Spin Glasses}},
  \href{https://doi.org/10.1007/s11040-014-9164-3}{\emph{Math. Phys. Anal.
  Geom.} {\bfseries 17} (2014) 441}
  [\href{https://arxiv.org/abs/1407.1552}{{\ttfamily 1407.1552}}].

\bibitem{feng2018spectrum}
R.~Feng, G.~Tian and D.~Wei, \emph{{Spectrum of SYK model}},
  \href{https://arxiv.org/abs/1801.10073}{{\ttfamily 1801.10073}}.

\bibitem{Micha2018}
M.~Berkooz, P.~Narayan and J.~Simon, \emph{{Chord diagrams, exact correlators
  in spin glasses and black hole bulk reconstruction}},
  \href{https://doi.org/10.1007/JHEP08(2018)192}{\emph{JHEP} {\bfseries 08}
  (2018) 192} [\href{https://arxiv.org/abs/1806.04380}{{\ttfamily
  1806.04380}}].

\bibitem{Berkooz_2019}
M.~Berkooz, M.~Isachenkov, V.~Narovlansky and G.~Torrents, \emph{{Towards a
  full solution of the large N double-scaled SYK model}},
  \href{https://doi.org/10.1007/JHEP03(2019)079}{\emph{JHEP} {\bfseries 03}
  (2019) 079} [\href{https://arxiv.org/abs/1811.02584}{{\ttfamily
  1811.02584}}].

\bibitem{Br_zin_1997}
E.~Brézin and S.~Hikami, \emph{Spectral form factor in a random matrix
  theory}, \href{https://doi.org/10.1103/physreve.55.4067}{\emph{Physical
  Review E} {\bfseries 55} (1997) 4067–4083}.

\bibitem{saad2018}
P.~Saad, S.~H. Shenker and D.~Stanford, \emph{{A semiclassical ramp in SYK and
  in gravity}},  \href{https://arxiv.org/abs/1806.06840}{{\ttfamily
  1806.06840}}.

\bibitem{Liu_2018}
J.~Liu, \emph{Spectral form factors and late time quantum chaos},
  \href{https://doi.org/10.1103/physrevd.98.086026}{\emph{Physical Review D}
  {\bfseries 98} (2018) }.

\bibitem{del_Campo_2017}
A.~del Campo, J.~Molina-Vilaplana and J.~Sonner, \emph{Scrambling the spectral
  form factor: Unitarity constraints and exact results},
  \href{https://doi.org/10.1103/physrevd.95.126008}{\emph{Physical Review D}
  {\bfseries 95} (2017) }.

\bibitem{winer2021hydrodynamic}
M.~Winer and B.~Swingle, \emph{Hydrodynamic theory of the connected spectral
  form factor},  2021.

\bibitem{Saad:2019lba}
P.~Saad, S.~H. Shenker and D.~Stanford, \emph{{JT gravity as a matrix
  integral}},  \href{https://arxiv.org/abs/1903.11115}{{\ttfamily 1903.11115}}.

\bibitem{Garc_a_Garc_a_2016}
A.~M. García-García and J.~J.~M. Verbaarschot, \emph{{Spectral and
  thermodynamic properties of the Sachdev-Ye-Kitaev model}},
  \href{https://doi.org/10.1103/PhysRevD.94.126010}{\emph{Phys. Rev.}
  {\bfseries D94} (2016) 126010}
  [\href{https://arxiv.org/abs/1610.03816}{{\ttfamily 1610.03816}}].

\bibitem{feng2018spectrum2}
R.~Feng, G.~Tian and D.~Wei, \emph{{Spectrum of SYK model II: Central limit
  theorem}},  \href{https://arxiv.org/abs/1806.05714}{{\ttfamily 1806.05714}}.

\bibitem{Altland_2018}
A.~Altland and D.~Bagrets, \emph{{Quantum ergodicity in the SYK model}},
  \href{https://doi.org/10.1016/j.nuclphysb.2018.02.015}{\emph{Nucl. Phys.}
  {\bfseries B930} (2018) 45}
  [\href{https://arxiv.org/abs/1712.05073}{{\ttfamily 1712.05073}}].

\bibitem{verbaarschot2019}
Y.~Jia and J.~J.~M. Verbaarschot, \emph{Spectral fluctuations in the
  sachdev-ye-kitaev model},
  \href{https://doi.org/10.1007/jhep07(2020)193}{\emph{Journal of High Energy
  Physics} {\bfseries 2020} (2020) }.

\bibitem{untajs_2020}
J.~{\v{S} }untajs, J.~Bon{\v{c}}a, T.~Prosen and L.~Vidmar, \emph{Quantum chaos
  challenges many-body localization},
  \href{https://doi.org/10.1103/physreve.102.062144}{\emph{Physical Review E}
  {\bfseries 102} (2020) }.

\bibitem{Aharony:2001pa}
O.~Aharony, M.~Berkooz and E.~Silverstein, \emph{{Multiple trace operators and
  nonlocal string theories}},
  \href{https://doi.org/10.1088/1126-6708/2001/08/006}{\emph{JHEP} {\bfseries
  08} (2001) 006} [\href{https://arxiv.org/abs/hep-th/0105309}{{\ttfamily
  hep-th/0105309}}].

\bibitem{Witten:2001ua}
E.~Witten, \emph{{Multitrace operators, boundary conditions, and AdS / CFT
  correspondence}},  \href{https://arxiv.org/abs/hep-th/0112258}{{\ttfamily
  hep-th/0112258}}.

\bibitem{Berkooz:2002ug}
M.~Berkooz, A.~Sever and A.~Shomer, \emph{{'Double trace' deformations,
  boundary conditions and space-time singularities}},
  \href{https://doi.org/10.1088/1126-6708/2002/05/034}{\emph{JHEP} {\bfseries
  05} (2002) 034} [\href{https://arxiv.org/abs/hep-th/0112264}{{\ttfamily
  hep-th/0112264}}].

\bibitem{MaldacenaMaoz}
J.~Maldacena and L.~Maoz, \emph{Wormholes in ads},
  \href{https://doi.org/10.1088/1126-6708/2004/02/053}{\emph{Journal of High
  Energy Physics} {\bfseries 2004} (2004) 053–053}.

\bibitem{Balasubramanian:2022tpr}
V.~Balasubramanian, P.~Caputa, J.~Magan and Q.~Wu, \emph{{A new measure of
  quantum state complexity}},
  \href{https://arxiv.org/abs/2202.06957}{{\ttfamily 2202.06957}}.

\bibitem{kobrin2020manybody}
B.~Kobrin, Z.~Yang, G.~D. Kahanamoku-Meyer, C.~T. Olund, J.~E. Moore,
  D.~Stanford et~al., \emph{Many-body chaos in the sachdev-ye-kitaev model},
  \href{https://doi.org/10.1103/physrevlett.126.030602}{\emph{Physical Review
  Letters} {\bfseries 126} (2021) }.

\end{thebibliography}\endgroup

\end{document}